\documentclass{PoS}
\pdfoutput=1

\title{$J/\psi$ production as a function of multiplicity in pp and p-Pb collisions with ALICE}

\ShortTitle{$J/\psi$ production as a function of multiplicity in pp and p-Pb collisions with ALICE}

\author{\speaker{Jana Crkovsk\'{a}}%
        \footnote{on behalf of the ALICE Collaboration}\\
       Institut de Physique Nucl\'{e}aire d'Orsay, Universit\'{e} Paris-Sud, CNRS-IN2P3, Orsay, France\\
       E-mail: \email{jana.crkovska@cern.ch}}

\abstract{
The multiplicity dependence of charmed-particle production can unveil new information on processes  taking part at the parton level and on the interplay of soft and hard production mechanisms in collisions of relativistic hadrons. 
In this contribution, we report on multiplicity-differential measurements of $\Jpsi$ in pp and p-Pb collisions studied by the ALICE Collaboration. 
Comparisons between measurements at different energies are drawn as well as comparisons with $\rm{D}$ mesons.
We also discuss the comparison with different theoretical predictions.
}

\FullConference{The European Physical Society Conference on High Energy Physics\\
         5-12 July\\
         Venice, Italy}

\newcommand{\dif}{\textup d}

\newcommand{\deriv}[2]{{\dif #1}/{\dif #2}}
\def\tup{\textup}
\newcommand{\Jpsi}{\mathrm{J}/\psi}

\newcommand{\snn}[2]{\sqrt{s_{\rm{NN}} } = #1 \tup{ #2}}
\newcommand{\spp}[2]{\sqrt{s} = #1 \tup{ #2}}

\newcommand{\Tm}{\ \textup{T} \cdot \textup{m}}

\newcommand{\abs}[1]{\vert #1 \vert}

\begin{document}

\section{Introduction}

Production of charmonia in collisions of ultrarelativistic hadrons is a subject of much scrutiny in high energy physics experiments. It combines hard processes, i.~e. the production of the initial charm quark-antiquark pair that can be described by means of perturbative Quantum ChromoDynamics (pQCD), with soft processes such as the subsequent hadronisation into a bound state. 
Moreover, particularly in case of $\Jpsi$, a non-negligible fraction of all produced particles originates from decays of $\rm{B}$-hadrons and its production is thus related to the production of bottom pairs in the collision \cite{nonprompt}. 
Production cross sections of charmonia and their dependence on the energy of the pp collision have been studied in ALICE \cite{Acharya:2017hjh}. The left panel of Fig. \ref{fig:Jpsi_ppXSec_and_7mult} shows the comparison of the forward $\Jpsi$ cross section measured in pp collisions at $\spp{13}{TeV}$ with theory.
$\Jpsi$ production is well described by the Non-Relativistic Quantum ChromoDynamics (NRQCD) formalism \cite{Ma:2010yw}. At forward rapidity, the results of the Colour Glass Condensate model (CGC) are added at low $p_{\rm{T}}$ to account for the gluon saturation at small $x$ in the incoming protons \cite{Ma:2014mri}. The so-called non-prompt contribution from $\rm{B}$ feed-down is well described by FONLL calculations \cite{Cacciari:2012ny}.
%

\section{Experimental setup}
ALICE \cite{Aamodt:2008zz} measures charmonia in two rapidity regions. 
At midrapidity $\abs{y_{\rm{lab}}}<0.9$, the dielectron decay channel is studied. 
The primary vertex is reconstructed by the Inner Tracking Systems (ITS) and the particle identification is provided by the Time Projection Chamber (TPC) via a specific energy loss measurement. 
The dielectrons are reconstructed from combined information from ITS and TPC.
The detectors are surrounded by a magent that generates a magnetic field of $B = 0.5 \textup{ T}$ along the beam direction.
Two scintillator arrays V0 located at  $-3.7 < \eta < -1.7$ and at $2.8 < \eta < 5.1$ are used for triggering purposes.
In the case of pp collisions, a large deposited charge in both V0 detectors will initiate a dedicated high-multiplicity trigger.
The multiplicity of charged particles at midrapidity $\deriv{N_{\rm{ch}}}{\eta}$ is proportional to the number of tracklets reconstructed by the Silicon Pixel Detector (SPD), which forms the two innermost layers of the ITS.

At forward rapidity $2.5 < y < 4$, the charmonia are reconstructed via their dimuon decay. The muon candidates are identified by the Muon Spectrometer. 
Muon candidates are detected and reconstructed in the Muon Chambers (MCH), consisting of 10 layers of Cathode Pad Chambers coupled to a $3 \Tm$ dipole magnet. 
A dedicated trigger on events containing muons is assured by a set of 4 planes of Resistive Plate Chambers grouped into 2 Muon Trigger Chambers (MTR).
Both MCH and MTR are shielded by a set of absorbers to assure a high-purity muon sample.

\section{Multiplicity-differential measurement of $\Jpsi$ in pp and p--Pb collisions}

Multiplicity-dependent studies of heavy-quark production in small systems can bring insight into the correlation between the soft particle-production mechanisms and hard processes occurring in hadron collisions.

\begin{figure}[h!]
    
    \begin{center}
     \includegraphics[width=.49\textwidth]{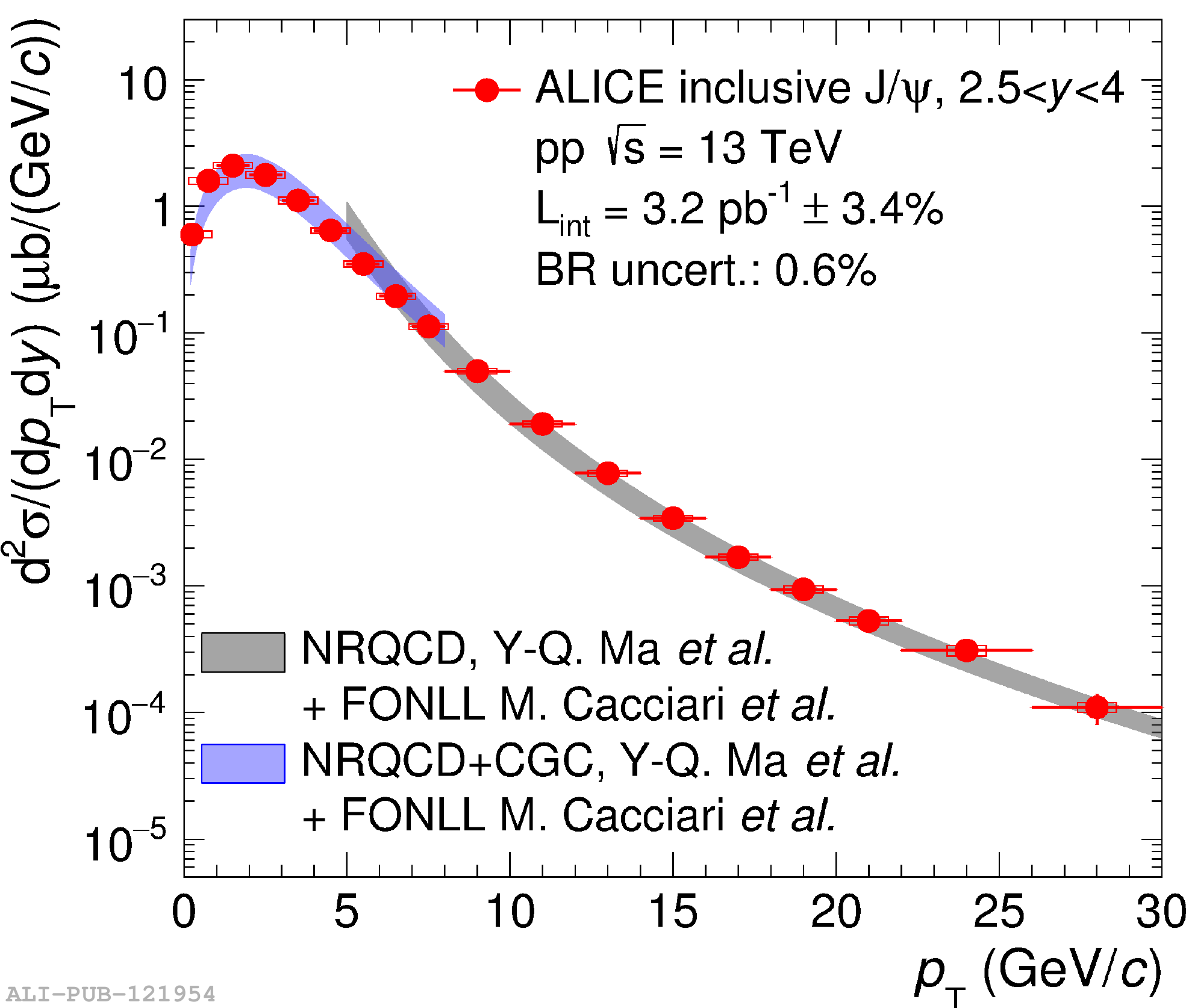}
     \includegraphics[width=.4\textwidth]{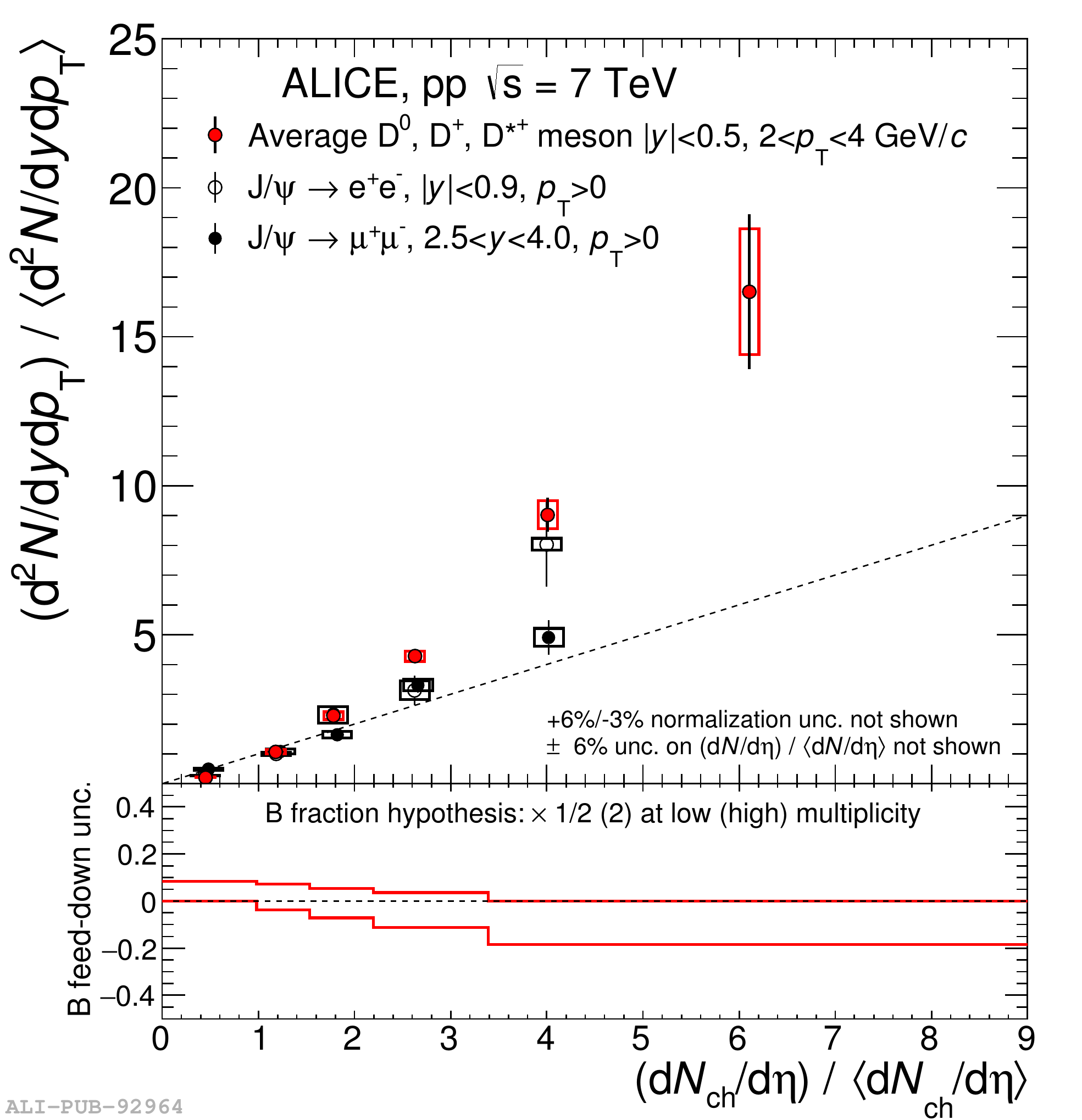}
      \caption{(Left) $\Jpsi$ cross section in pp collisions at $\spp{13}{TeV}$ \cite{Acharya:2017hjh}, compared with a theoretical prediction for the prompt contribution \cite{Ma:2010yw,Ma:2014mri} summed with results of FONLL calculations \cite{Cacciari:2012ny}. (Right) Multiplicity dependence of relative $\Jpsi$ and $\rm{D}$ meson yields in pp at $\spp{7}{TeV}$ \cite{Adam:2015ota}. The bottom panel shows the uncertainty on the feed-down fraction for the $\rm{D}$ mesons. }
      \label{fig:Jpsi_ppXSec_and_7mult}
    \end{center}

  \end{figure}

ALICE measured the multiplicity dependence of mid- and forward-rapidity $\Jpsi$ in pp collisions at $\spp{7}{TeV}$ \cite{Abelev:2012rz}. The $\Jpsi$ yields show an increase with multiplicity which is consistent with the one observed for measurements of $\rm{D}$ mesons at the same energy \cite{Adam:2015ota}; the comparison is shown in Fig. \ref{fig:Jpsi_ppXSec_and_7mult} right.  
While the $\rm{D}$ mesons show a stronger-than-linear increase, the large uncertainties on the $\Jpsi$ measurement prevent the determination of the nature of the dependence. Yet the similar behaviour of the two species hints that the observed effect is independent of hadronisation and is instead likely to be linked to the creation of the charm quark-antiquark pair.

\begin{figure}[h!]
  
    \begin{center}
     \includegraphics[width=.49\textwidth]{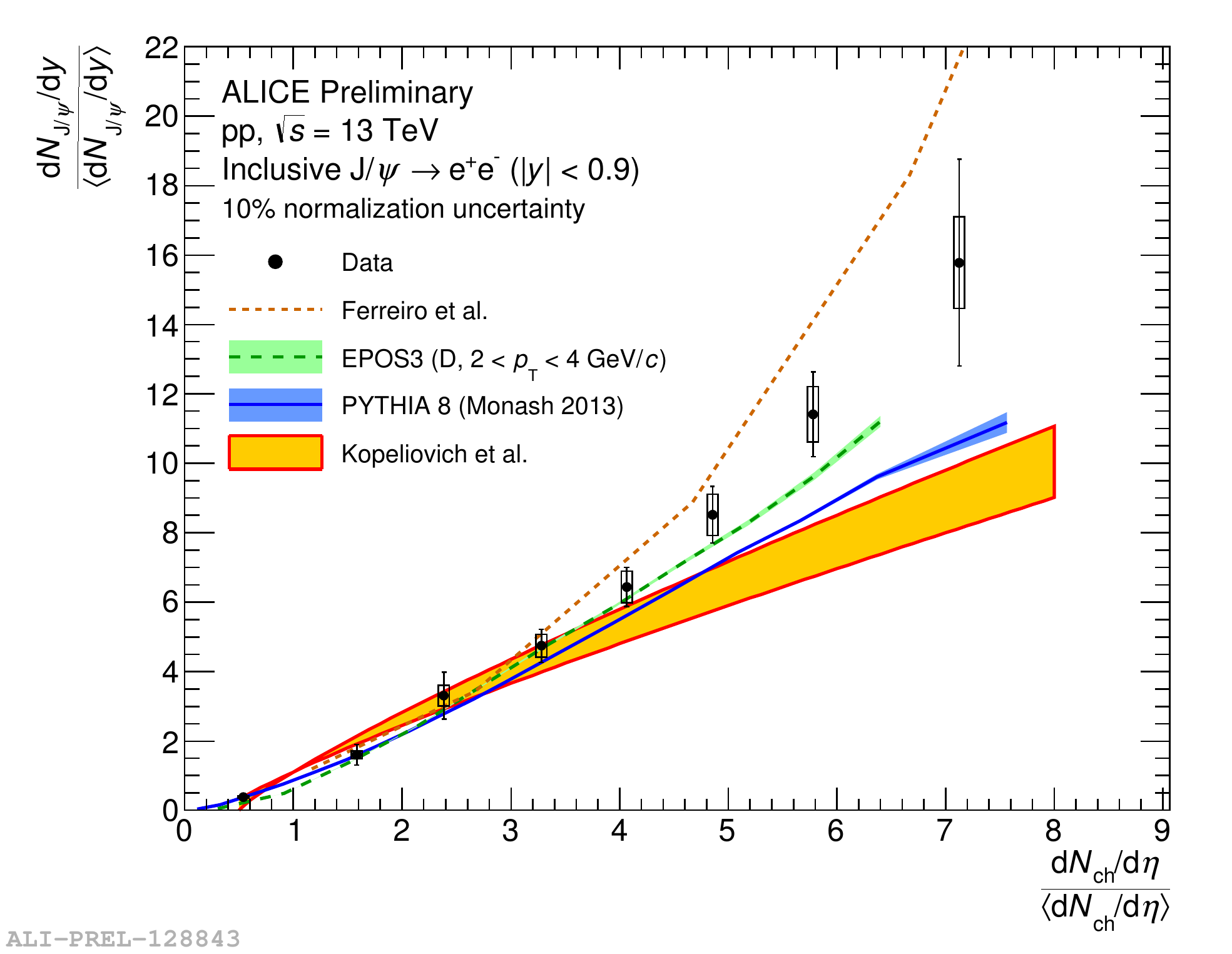}
     \includegraphics[width=.49\textwidth]{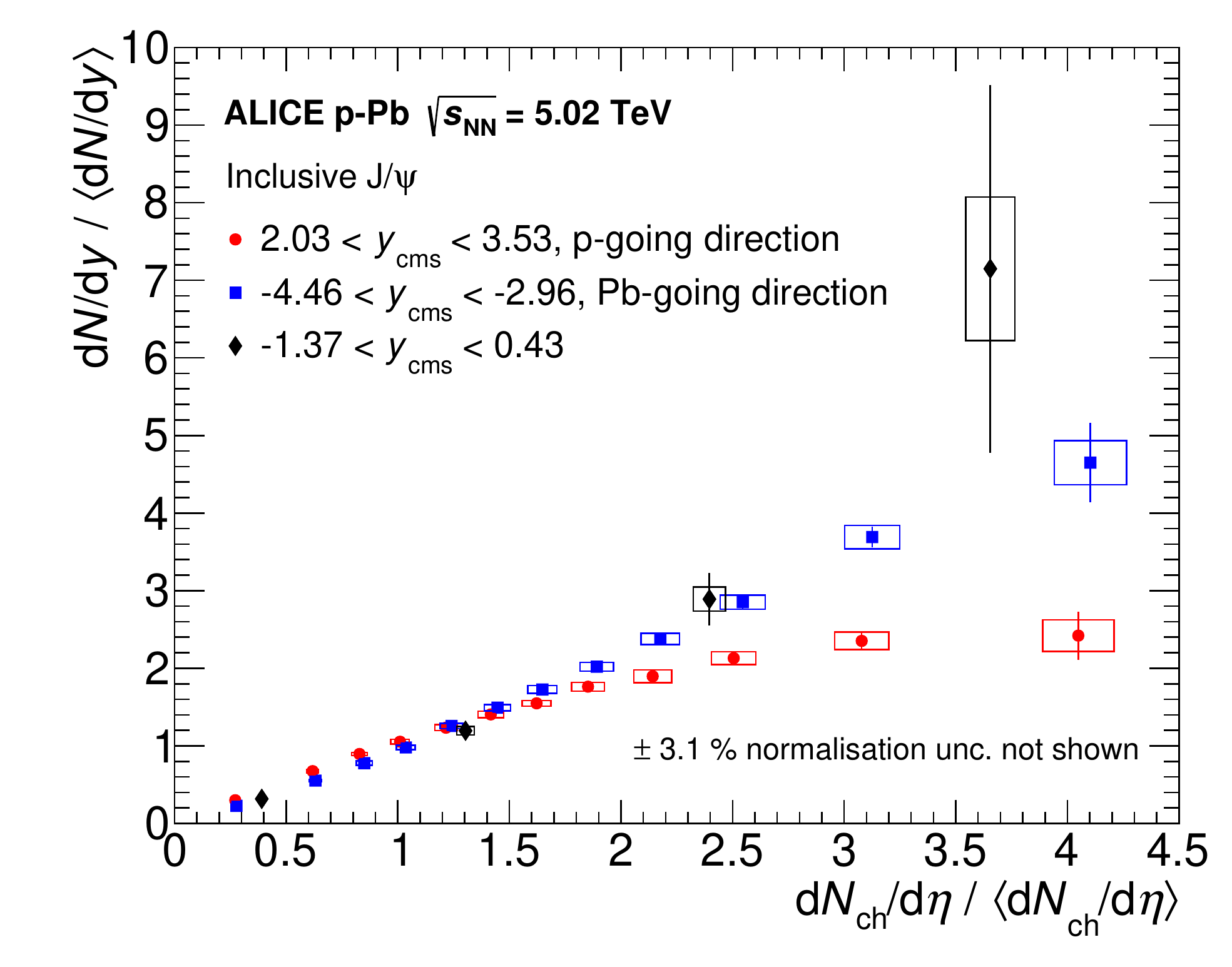}
      \caption{(Left) Multiplicity dependence of the relatve inclusive $\Jpsi$ yields at midrapidity in pp collisions at $\spp{13}{TeV}$ \cite{Weber:2017hhm} compared with predictions from Ferreiro \cite{Ferreiro:2012fb}, EPOS 3 \cite{epos3}, PYTHIA 8 \cite{Sjostrand:2007gs}, and Kopeliovich \cite{Kopeliovich:2013yfa}. (Right) Multiplicity dependence of relative $\Jpsi$ yields in p--Pb at mid-, forward and backward rapidity \cite{Adamova:2017uhu}. }
      \label{fig:Jpsi_pp13_and_pPb5}
    \end{center}

  \end{figure}

Recently, ALICE has measured the multiplicity dependence of midrapidity $\Jpsi$ at $\spp{13}{TeV}$ \cite{Weber:2017hhm}. Due to the large statistics of the 13 TeV data sample and the use of the dedicated high-multiplicity trigger, the reach in multiplicity is nearly doubled with respect to the 7 TeV measurement, which is sufficient to discern the nature of the dependence.
The yields increase in a stronger-than-linear fashion as can be seen in the left panel of Fig. \ref{fig:Jpsi_pp13_and_pPb5}. The data are compared with available theoretical predictions \cite{Ferreiro:2012fb,epos3,Sjostrand:2007gs,Kopeliovich:2013yfa}. 
The models used in this comparison all implement, each of them in a different way, multiple-parton-interaction (MPI) processes, which result in an increase of yields with multiplicity. The nature of the increase differs between individual models. All shown models provide a qualitative description of the measured yields within uncertainty, however at current precision the data favour none of the MPI scenarios.
 
Measurements of $\Jpsi$ production in p--Pb collisions allow the study of effects caused by the presence of the nuclear medium in the collision, the so-called cold nuclear matter effects. ALICE can probe three distinct rapidity regions in p--Pb collisions due to two running periods with inverted beam directions.
The measurement at forward rapidity, i.~e. when the proton comes towards the spectrometer, probes the small $x$ of the incoming Pb nucleus where cold nuclear matter effects such as gluon shadowing and saturation effects are expected.

A measurement of multiplicity-dependent $\Jpsi$ yields in p--Pb collisions at $\snn{5.02}{TeV}$ was caried out\cite{Adamova:2017uhu}.
The results for the three rapidity regions can be found in the right panel of Fig. \ref{fig:Jpsi_pp13_and_pPb5}.
The midrapidity ($-1.37 < y_{\rm{cms}} < 0.43$) and backward-rapidity $\Jpsi$ yields ($-4.46 < y_{\rm{cms}} < -2.96$) exhibit a linear increase with the multiplicity of the event. 
On th other hand, the forward-rapidity yields ($2.03 < y_{\rm{cms}} < 3.53$) saturate at multiplicity above $1.5-2$ times the average minimum-bias multiplicity.

Moreover, the midrapidity measurement was compared with prompt $\rm{D}$-meson data in the same collision system \cite{Adam:2016mkz}. The behaviour of the two species is similar, as observed earlier in pp collisions at $\spp{7}{TeV}$ \cite{Abelev:2012rz}.

\section{Summary}

We reviewed multiplicity-differential measurements of $\Jpsi$  in pp and p--Pb collisions in ALICE. 
In pp collisions, $\Jpsi$ were measured as a function of multiplicity at $\spp{7 \mbox{ and } 13}{TeV}$. The midrapidity yields show an increase with multiplicity, which is consistent with the one observed for midrapidity $\rm{D}$ mesons. 
The measurement at midrapidity at $\spp{13}{TeV}$ \cite{Weber:2017hhm} shows a behaviour consistent with the 7 TeV data. Due to larger statistics, the on-going measurement at 13 TeV allows us to double the measured multiplicity region and thus to better determine the nature of the increase. The recent data are qualitatively described by theoretical models considering MPI processes. 

Multiplicity-differential measurements of $\Jpsi$ were also studied in p--Pb collisions at $\snn{5.02}{TeV}$ in three different rapidity regions. While the midrapidity and forward-rapidity yields show a linear increase, the backward measurements displays a saturation of yields. 
Newly available p--Pb data at $\snn{8.16}{TeV}$ should extend the measurement to higher multiplicities and provide more precise data to reveal the nature of the increase at different rapidities.

\end{document}